\documentclass[sigconf,preprint,nonacm]{acmart}

\usepackage{booktabs}
\usepackage{float}
\usepackage{subcaption}

\AtBeginDocument{%
  }

\setcopyright{acmlicensed}
\copyrightyear{2018}
\acmYear{2018}
\acmDOI{XXXXXXX.XXXXXXX}
\acmConference[SIGIR 2026]{SIGIR 2026 Submission}{July 2026}{Melbourne, Australia}
\acmISBN{978-1-4503-XXXX-X/2018/06}

\usepackage{hyperref}
\begin{document}

\title{IncompeBench: A Permissively Licensed, Fine-Grained Benchmark for Music Information Retrieval}

\author{Benjamin Clavié}
\email{ben@mixedbread.com}
\affiliation{%
  \institution{Mixedbread AI and National Institute of Informatics (NII)}
  \city{Tokyo}
  \country{Japan}
}

\author{Atoof Shakir}
\email{atoof@mixedbread.com}
\affiliation{%
  \institution{ETHZ and Mixedbread AI}
  \city{Zürich}
  \country{Switzerland}
}

\author{Jonah Turner}
\author{Sean Lee}
\author{Aamir Shakir}
\affiliation{%
  \institution{Mixedbread AI}
  \city{San Francisco}
  \state{CA}
  \country{USA}
}

\author{Makoto P. Kato}
\affiliation{%
  \institution{University of Tsukuba and NII}
  \city{Tsukuba}
  \country{Japan}
}

\renewcommand{\shortauthors}{Clavié et al.}

\begin{abstract}
  Multimodal Information Retrieval has made significant progress in recent years, leveraging the increasingly strong multimodal abilities of deep pre-trained models to represent information across modalities. 
  Music Information Retrieval (MIR), in particular, has considerably increased in quality, with neural representations of music even making its way into everyday life products.
  However, there is a lack of high-quality  benchmarks for evaluating music retrieval performance.
  To address this issue, we introduce \textbf{IncompeBench}, a carefully annotated benchmark comprising $1,574$ permissively licensed, high-quality music snippets, $500$ diverse queries, and over $125,000$ individual relevance judgements.
  These annotations were created through the use of a multi-stage pipeline, resulting in high agreement between human annotators and the generated data.

\end{abstract}




\maketitle




\section{Introduction}

The ability to search for and retrieve relevant music from large collections is one of the important challenges on the way to better multimodal information retrieval. With the democratisation of music production, it is estimated that the music production of the last ten years vastly exceeds the quantity of music that had been created in all preceding decades, with no suggestion of this trend slowing down~\cite{musictrends}.

Additionally, as semantic search has become more commonplace, users increasingly rely on natural ways of querying large collections. For music, this manifests as queries ranging from terse keyword searches and conversational queries to descriptions of mood, genre, ``vibe'', instrumentation, tempo, suitability as background music for a given situation, etc... to find music that matches their intent. 
These factors have driven an increasing volume of interest towards neural approaches in text-to-audio Music Information Retrieval (MIR), where recent advances in multimodal representation learning have led to increasingly capable retrieval models~\cite{clamp3,clap,ttmr}.

Despite these advances, progress in music retrieval is difficult to measure reliably. The licensing issue has historically been a persistent obstacle. 
The vast majority of commercially recorded music is protected by copyright, making it difficult to distribute audio corpora alongside benchmark annotations. 
In practice, this has led researchers to either rely on proprietary datasets that cannot be shared~\cite{universal2}, or on lower-quality or synthetic datasets of limited realism.

Furthermore, evaluation in information retrieval has long depended on high-quality, publicly available benchmarks with dense relevance judgements, and the construction of such resources has been central to advances in text retrieval, from the TREC ad-hoc tracks~\cite{dl19,dl20} to MS~MARCO~\cite{msmarco} and BEIR~\cite{beir}. These benchmarks share key properties: diverse, good quality queries, graded or dense relevance annotations, and sufficient signal to distinguish meaningfully between systems. Existing music retrieval benchmarks, however, often fall short on one or more of these dimensions. Many rely on binary relevance or simple tag-matching, failing to capture the inherently graded nature of music similarity, where a query for \emph{``upbeat jazz with piano''} may be partially satisfied by a bossa nova piano track, well-matched by a swinging jazz trio, and perfectly answered by a high-tempo bebop piece. Others are constrained by small scale, narrow query diversity, or restrictive licensing that limits reproducibility and redistribution.

In this work, we introduce \textbf{IncompeBench}, a fine-grained music audio retrieval benchmark designed to address these shortcomings and support high-quality, fine-grained MIR evaluations. IncompeBench is built on the IncompeTech music collection, comprising over 1{,}500 high-quality tracks released under a permissive license, providing a diverse, shareable audio corpus spanning a wide range of genres, moods, and instrumentation.

\textbf{IncompeBench} is composed of 500 queries with controlled variation in style (keywords, questions, descriptions, conversational, and imperative), length, attribute complexity, and negation, attempting to capture at least some aspects of real users' music search behaviour, with 128,000 graded relevance annotations. To support this, we introduce a multi-stage automated pipeline to generate fine-grained annotations at scale. We first create detailed \emph{song cards} summarising each track's musical attributes using a frontier multimodal model, before using these song cards to seed our query generation step. We then select annotation candidates through a model-diverse retrieval and fusion strategy, and perform pointwise relevance labelling with Gemini~3~Pro using a prompt adapted from the UMBRELA relevance annotation framework~\cite{UMBRELA} used by TREC-RAG~\cite{trecrag}. The resulting annotations use a four-level graded relevance scale (0--3) with over 125{,}000 individual relevance judgements. A human verification study with expert annotators yields a quadratic weighted Cohen's $\kappa$ of 0.94, indicating strong alignment between automated and human assessments.

Informed by this agreement analysis, which reveals that the primary source of annotation noise is the tendency of LLM annotators to be overly lenient on what constitutes partial relevance, we release two evaluation variants: \textbf{IncompeBench-Lenient}, retaining all three positive relevance levels, and \textbf{IncompeBench-Strict}, which discards tangential annotations and retains only clearly relevant judgements. Finally, we provide baseline results for several current music retrieval models across both settings, showcasing both overall low performance and meaningful differences between existing models, further validating the usefulness of IncompeBench as a measure of music retrieval performance.

We publicly release both variants of IncompeBench, including songs, queries and annotated qrels, as well as the prompts and DSPy programs used to generate publicly, under the original CC-BY license, respecting the original song corpus licensing.

\section{Related Works}

\paragraph{Music-Language Datasets and Benchmarks.}
The development of benchmarks connecting music and natural language has accelerated in recent years~\cite{audiotrends}, in line with the trends towards increasingly multimodal models~\cite{multimodaltrends}. MusicCaps, released alongside MusicLM~\cite{musiclm}, provides 5.5k expert-annotated audio-caption pairs, and has become a standard evaluation resource for music captioning and retrieval. However, its clips are sourced from YouTube under restrictive terms, and the dataset provides only single captions per clip with no graded relevance annotations, making it unsuitable for fine-grained retrieval evaluation. The Song Describer Dataset (SDD)~\cite{songdescriber} addresses some licensing concerns by pairing 1.1k crowdsourced captions with 706 permissively licensed recordings, and has demonstrated the importance of cross-dataset evaluation for music-language models. While SDD represents a valuable step towards open, reproducible evaluation, it remains small in scale, offers only binary caption-audio matching, and is designed primarily for captioning and generation evaluation rather than for retrieval with graded relevance. LP-MusicCaps~\cite{lpmusiccaps} scales music captioning data to 2.2M pseudo-captions generated via LLMs, but these are of varying quality, lack strong human validation, and were generated through models now known to have limited fine-grained understanding abilities. Indeed, LP-MusicCaps is intended to serve as a large-scale training dataset rather than as a fine-grained evaluation set. Other large-scale collections, such as the Jamendo tagging dataset~\cite{jamendo}, exhibit similar limitations, largely relying on sparsely annotated corpora with captions. Beyond music-text datasets, the WikiMT-X benchmark released with CLaMP~3~\cite{clamp3} offers 1,000 triplets of sheet music, audio, and text descriptions, but evaluates paired retrieval in controlled settings (with captions) rather than ranked retrieval with natural language queries and graded annotations. In the broader IR community, the construction of high-quality benchmarks with dense relevance judgements has been central to advancing retrieval research. The TREC series~\cite{dl19,dl20}, MS~MARCO~\cite{msmarco}, and BEIR~\cite{beir} have driven progress in text retrieval through large-scale, graded, and diverse annotations. Recent domain-specific benchmarks such as FollowIR~\cite{followir} and ToolRet~\cite{toolret} demonstrate continued demand for targeted evaluation resources in underserved retrieval domains, using automated pipelines validated by human annotators to construct challenging benchmarks at scale.

%
\paragraph{Text-to-Music Retrieval Models.}
Neural approaches to text-to-music retrieval have progressed rapidly. CLAP~\cite{clap} represented a landmark improvement, inspired by the vision-specific CLIP~\cite{clip}, by adapting contrastive language-audio pretraining to align audio and text in a shared embedding space through training on large-scale general audio datasets. Since then, TTMR++~\cite{ttmr} improves upon earlier text-to-music retrieval models by enriching training descriptions with LLM-generated captions and artist metadata, and evaluates on MusicCaps and SDD using Recall@10. CLaMP~3~\cite{clamp3}, the current state of the art on most existing tasks, extends contrastive learning to align all major music modalities---sheet music, MIDI, and audio---with multilingual text, training on 2.31M music-text pairs. Industrial efforts have also produced strong retrieval systems: MULE~\cite{mule}, trained on large-scale expert-annotated proprietary music data at Pandora/SiriusXM, demonstrates that supervised pre-training on industry catalogues yields powerful audio representations, though neither the training data nor the evaluation sets used can be publicly shared. This pattern of reliance on proprietary audio catalogues and internal evaluation sets is common across the industry, with multiple recent, highly performing audio retrievers such as SLAP~\cite{universal2} and GD-Retriever~\cite{universal} describing strong models with proprietary training and evaluations, limiting reproducibility and making it difficult to compare systems on common ground. Meanwhile, recent work such as ColQwen-Omni~\cite{colqwenomni} represents an attempt to transfer the late-interaction retrieval recipe of ColPali~\cite{colpali} from document retrieval to music through omnimodal base models. Across all of these models, evaluation for retrieval has been limited to either paired retrieval metrics on small caption datasets (MusicCaps, SDD) or proprietary benchmarks, further reinforcing the need for additional high-quality evaluation datasets.

\section{Benchmark Building}

\subsection{The Song Corpus: Choosing IncompeTech}

We first focused on selecting a suitable source of song snippets. Our search focused on a set of constraints: the songs had to be of high audio quality, permissively licensed to support open research use without risks of copyright infringement, diverse enough to cover a broad range of styles, genres and instruments, and in small enough volume to make fine-grained annotation for each query possible.

The only collection of songs which we find to meet all of these constraints without significant processing efforts and licensing constraint is IncompeTech\footnote{\href{https://incompetech.com/}{https://incompetech.com/}}, a famous collection of over 2,000 songs composed, recorded and released under a CC-BY license by Kevin MacLeod\footnote{While the full IncompeTech collection can be downloaded song-by-song, we chose to make a donation in order to receive bulk download access.}. As a result of its high quality, broad genre coverage and permissive licensing, the IncompeTech collection has achieved a significant foothold in popular culture, being among the most listened-to song collections in the world according to numerous media outlets~\cite{nyt1,mostlistened2}. 
While not a criterion in itself, we believe this increases the likelihood of it being a strong proxy for real-world music retrieval.

\subsection{Corpus Preparation}

The full IncompeTech dataset, at the time of downloading it, contained over 2,000 songs. As part of our pre-preprocessing, we first excluded all songs shorter than 90 seconds and then performed chunk generation through a simple logic: for each of the remaining songs, we created three 30 second chunks, one starting from the beginning of the song, and the other two starting at randomly determined points. The chunks are sampled at an audio rate of 16kHz, maintaining quality while facilitating processing.

Unlike in text retrieval, where chunks of a single document have varying degrees of relevance to a given query~\cite{msmarco,miracl}, the vast majority of chunks extracted from a given song shared identical attributes, meaning they would very frequently all be high-quality matches for a given query. Therefore, to keep the annotation process tractable and to avoid spending significant budget on judging chunks from the same songs, we randomly sample just one of these three chunks per song to create the final song corpus.

The resulting corpus, which we use in the rest of this paper, is composed of 1574 individual 30 seconds audio chunks.

\subsection{Query Generation}

Unless otherwise specified, the DSPy programs for the steps detailed below are made publicly available~\footnote{\href{https://github.com/mixedbread-ai/incompebench-programs}{https://github.com/mixedbread-ai/incompebench-programs}}.

\subsubsection{Creating Song Cards}
\label{sec:songcard}

During early experimentation, we found that frontier models such as Gemini 3 Pro were reliably able to extract information to a good degree of specificity, for example distinguishing between specific, obscure types of string instruments, or accurately identifying sub-genres of Latin music. However, this information appeared lost when attempting to prompt the model for query generation directly, with the queries instead focusing on superficial elements. Additional prompting for reasoning did not seem to noticeably improve this behaviour.

We therefore modified our pipeline to work as a two-stage process, where each song is first carefully analysed by the model to produce a "song card", capturing varied elements such as rhythm, tempo, genres, artist-soundalike/inspiration, instruments, etc... In total, around 30 attributes were identified per song to be used as targeted attributes, as described in Section~\ref{sec:gen} below.

To generate the queries, both the song cards and the audio were passed to the query generation model. We found that this preliminary attribute action step greatly enhanced the diversity of attributes targeted by the queries, where direct prompting tended to collapse towards "caption"-style queries, as is common in existing audio benchmarks such as LP-MMCaps~\cite{lpmusiccaps} or the Song Describer Dataset~\cite{songdescriber}, possibly suggesting an over-reliance on previously seen data.

\begin{table*}[t]
\centering
\setlength{\tabcolsep}{4pt}
\renewcommand{\arraystretch}{0.95}
\begin{minipage}[t]{0.48\textwidth}
\centering
\begin{tabular}{lr}
\toprule
\textbf{Query characteristic} & \textbf{Proportion} \\
\midrule
Short keywords & $\approx$40\% \\
Question & $\approx$20\% \\
Descriptive & $\approx$25\% \\
Conversational & $\approx$15\% \\
\midrule
1 attribute & $\approx$40\% \\
2 attributes & $\approx$30\% \\
3 attributes & $\approx$20\% \\
4 attributes & $\approx$10\% \\
\midrule
Negation (overall) & $\approx$12\% \\
\bottomrule
\end{tabular}
\caption{Distribution of query characteristics across the generated queries.}
\label{tab:attributestats}
\end{minipage}
\hfill
\begin{minipage}[t]{0.48\textwidth}
\centering
\begin{tabular}{lrr}
\toprule
& \textbf{Words} & \textbf{Tokens} \\
\midrule
\multicolumn{3}{l}{\textbf{Query length}} \\
\midrule
Min & 3 & 3 \\
P25 & 6 & 7 \\
P50 & 8 & 10 \\
P75 & 12 & 15 \\
P90 & 21 & 24 \\
Max & 26 & 31 \\
\midrule
Mean & 10.3 & 12.3 \\
Std & 5.7 & 6.5 \\
\bottomrule
\end{tabular}
\caption{Length distribution of queries.}
\label{tab:qlength}
\end{minipage}
\label{tab:querystats}
\end{table*}

\subsubsection{Generation Step}
\label{sec:gen}

Following the creation of song cards for every song, we randomly sample 500 songs which are used as seed documents for the query generation phase.

Queries are generated with sets of constraints identified in the prompt. We follow Qwen3-Embeddings~\cite{qwen3embed} in adopting a two-stage query generation method. For both stages, outputs are generated using declarative ``program`` through the DSPy framework~\cite{dspy} rather than with plain-text prompting.

In the initial stage, the model is provided with the song card and the audio snippet, as well as with four potential user personas randomly sampled from NVidia's Nemotron Persona~\cite{nemopersona} dataset. In this first stage, the model is asked to select which persona would be likely to enquire about this specific song, and selects specific elements of the song that the query should target.

This information is then provided to the actual query generation stage. In addition to the persona and elements sampled during stage 1, specific constraints are introduced in order to induce query variety. These attributes are \textbf{Number of attributes}, how many attributes of the song, previously identified in their Song Cards, should be covered by the query (between 1 and 4); \textbf{Query style}, whether the query should be keyword-style, a question, an instruction, conversational in nature, or descriptive of the target song; \textbf{Query length}, how many words the query should contain; and \textbf{Negation rate}, whether or not one or more of the selected attributes should be negative rather than positive attributes (e.g. ``high BPM song without guitar'')

The choice of these attributes and the weight given to each possible value were empirically deduced through analysing a private set of real-world user queries gathered with user consent by a commercial search service. Further information on the resulting distribution of queries is provided in Section~\ref{sec:stats}. Additionally, to encourage query variety and reduce repeated phrasings, the model is prompted to generate two queries in each output, and one of them is randomly sampled while the other is discarded. Finally, we note that negations are only applied to queries with more than 1 attribute targeted, so as to avoid overly-broad, purely negative queries.

\subsection{Annotation Candidate Selection}

While our dataset contains a relatively small number of queries (500) and documents (1,574), validating every query-document relationship is impractical, as this would represent over 850000 individual annotations to be generated by a frontier LLM, followed by significant human validation efforts.

However, it remains important that many pairs are individually judged to allow for fine-grained evaluations, as music queries based on musical attributes can have a large number of potential matches. To strike a balance between these constraints, allowing for significant breadth of judgement while remaining tractable, we generate large lists of candidate songs for each individual query.

Our candidate generation pipeline includes multiple steps. First, we retrieve the top 500 candidates for each query, using many different retrieval models: CLAMP3~\cite{clamp3}, TTMR++, CLAP~\cite{clap}, ColQwen-Omni, as well as a proprietary internal music retrieval model which we have found to perform well on other musical retrieval tasks. It has previously been observed that due to the significant efforts and vast amounts of data used to train text retrieval models, they are often competitive with multi-modal retrieval methods in cases where the multimodal data can be accurately transcribed into text~\cite{miraclvision}. To take advantage of this and add more diversity to our data, we also use mixedbread-embed-large~\cite{mxbaiembed} to retrieve song cards, as described in Section~\ref{sec:songcard}, for each of the 500 generated queries.

Using all of these top-500 lists, we then perform Reciprocal Rank Fusion (RRF)~\cite{rrf} to generate top-250 lists for each query. In addition, to avoid giving undue advantage to certain baseline models in the future, we enrich the candidate list for each query with any result found in the top 30 for a given model that did not make it onto the final fused top 250 list. In total, this results in around 128,000 query <-> song pairs being annotated.

We believe that this diverse candidate generation step is a good way to evaluate retrieval and ranking quality under candidate generation constraints, similar to popular text collections such as MS MARCO~\cite{msmarco} and the TREC DL series~\cite{dl19,dl20}.

\subsection{Automated Labelling}

The labelling process is conducted with Gemini 3 Pro~\cite{gemini3}, which we found to currently be the model able to most consistently produce high-quality ratings that were well aligned with human reviews, as further detailed in Section~\ref{sec:agreement}. We chose to use a prompt inspired by UMBRELA~\cite{UMBRELA}, with fine-grained relevance levels, ranging from 0 (completely irrelevant) to 3 (fully relevant to every aspect of the query). The full prompt for this stage is available at \href{https://github.com/mixedbread-ai/incompebench-programs/tree/main}{this URL}.

The choice of Gemini 3 Pro was motivated by empirical findings. During our initial exploration, we initially experimented with Gemini 3 Flash~\cite{gemini3} and Qwen3-Omni-30B-A3B~\cite{qwen3omni}, two high-quality models with very high cost-efficiency. However, after an analysis of sample runs and despite prompt modifications, we found the models' reasoning and resulting annotations to be lacking. While they were both able to broadly capture high-level relevance, ensuring songs were broadly in the target genre or mood, both models consistently made errors in their assessment of finer details and frequently labeled songs as fully-relevant or completely irrelevant based on these missed details.

\begin{figure*}[htbp]
    \centering
    \begin{minipage}{0.85\textwidth}
        \centering
        \begin{subfigure}[t]{0.48\linewidth}
            \centering
            \includegraphics[width=\linewidth]{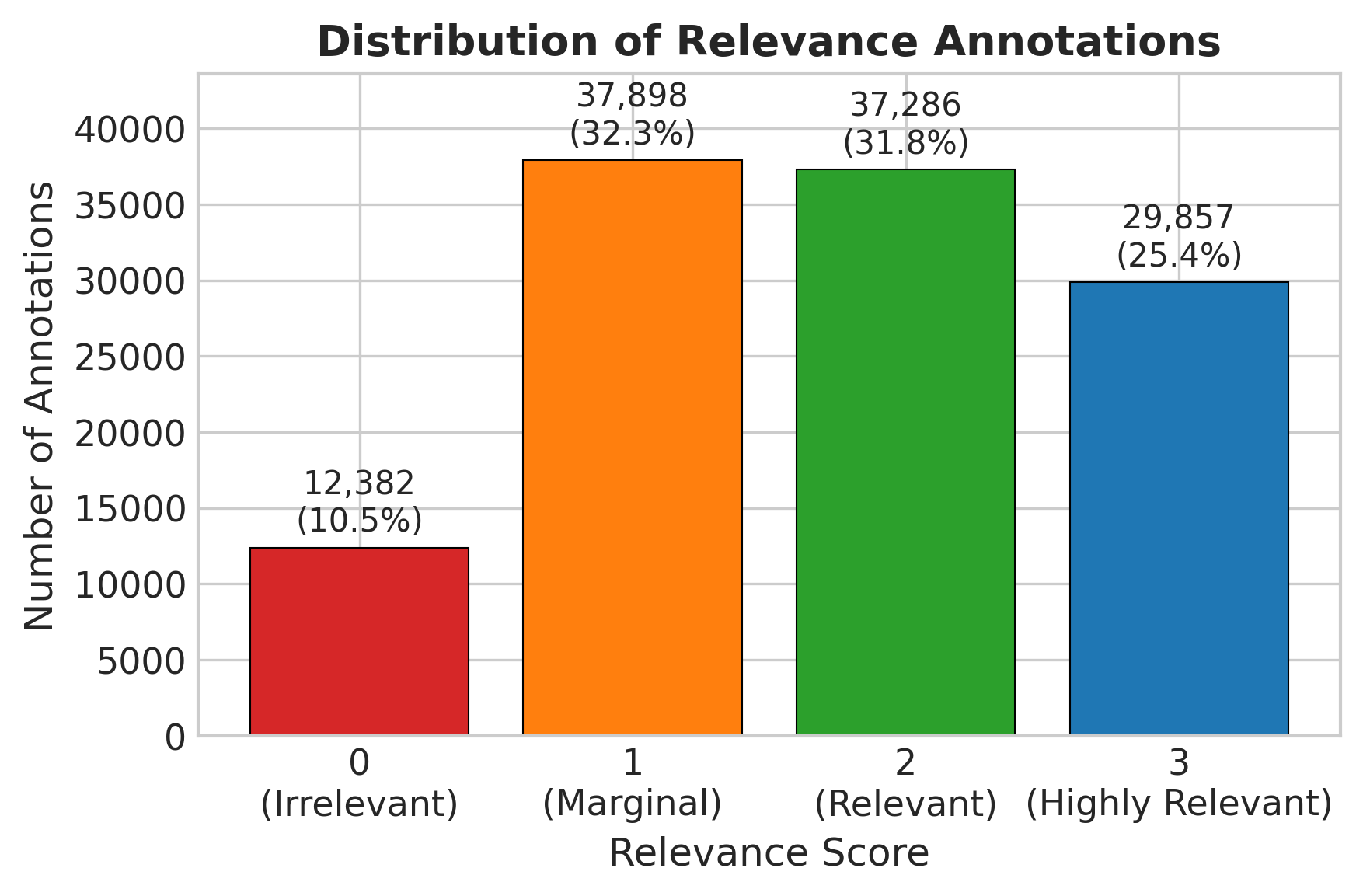}
            \caption{Overall annotation distribution}
            \label{fig:labeldistrib}
        \end{subfigure}
        \hfill
        \begin{subfigure}[t]{0.48\linewidth}
            \centering
            \includegraphics[width=\linewidth]{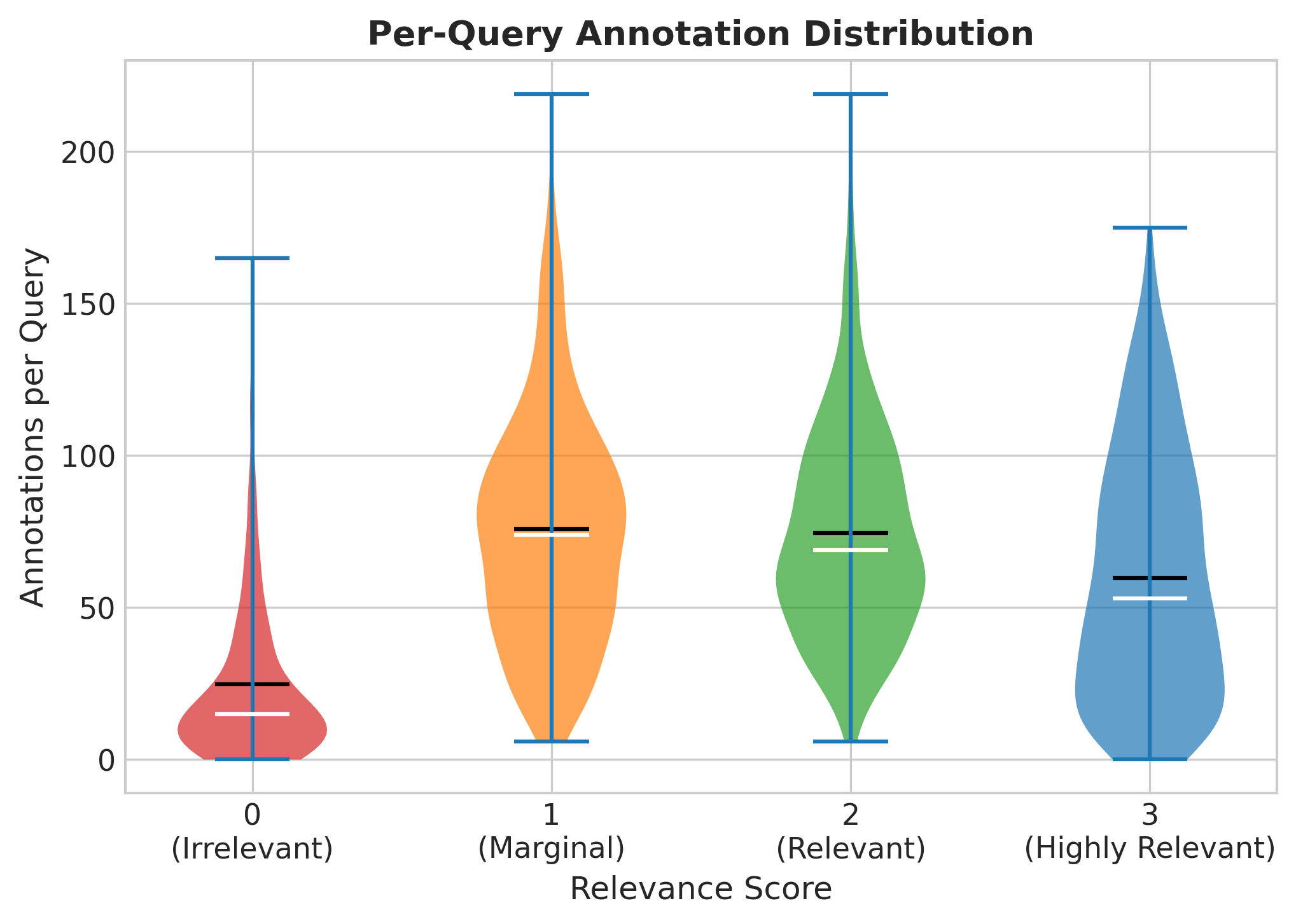}
            \caption{Per-query annotation distributions}
            \label{fig:perquerydistrib}
        \end{subfigure}
        \caption{Annotation distributions at the corpus and query level.}
        \label{fig:combined}
    \end{minipage}
\end{figure*}

\begin{table}[t]
\centering
\begin{tabular}{lcc}
  \toprule
  LLM-Assigned label & Precision & Re-labels on disagreement \\
  \midrule
  0 & 0.99 & 1 ($\approx$99\%) \\
  1 & 0.83 & 0 ($\approx$96\%), 2 ($\approx$3.5\%), 3 ($\approx$0.5\%) \\
  2 & 0.93 & 1 ($\approx$62\%), 3 ($\approx$37\%), 0 ($\approx$5\%) \\
  3 & 0.94 & 2 ($\approx$92\%), 1 ($\approx$7\%), 0 ($\approx$2\%) \\
  \midrule
  Macro-average & 0.92 & -- \\
  \midrule
  \midrule
  Quad.\ weighted $\kappa$ & \multicolumn{2}{c}{0.94} \\
  \bottomrule
\end{tabular}
\caption{Human verification of LLM relevance labels ($n=385$). Precision refers to the model's precision against the human assessment.}
\label{tab:agreement}
\end{table}
\section{IncompeBench}

In this section, we will present an analysis of the constructed benchmark and detailed statistics of the generated queries and annotations in Section~\ref{sec:stats}, followed by a study of the LLM-Expert Human agreement in Section~\ref{sec:agreement} and the definition of the final evaluation sets, informed by this study, in Section~\ref{sec:strict}.

\subsection{Benchmark Statistics}
\label{sec:stats}

In Table~\ref{tab:attributestats}, we present the high-level overall distribution of query styles, as previously described in Section~\ref{sec:gen}. We observe that the queries are phrased with a diverse style, with short keywords being the dominant individual query format, which is mitigated by being the only non-natural language style out of the 5. Table~\ref{tab:qlength} presents the distribution of the query phrasings, in both word and Qwen3~\cite{qwen3} --- the most popular open-source LLM family --- token counts, so as to provide an overview of the data make-up. Overall, the combination of these distributions matches results observed in a proprietary commercial search system, suggesting a good coverage of common query styles.

Finally, Figure~\ref{fig:labeldistrib} shows the distribution of the labels assigned by Gemini 3 Pro used to generate the qrels of our final benchmark, and Figure~\ref{fig:perquerydistrib} presents the distribution of positive qrels per query, showcasing the diversity of the corpus. The resulting distribution matches our initial expectations, as open-ended queries related to music are likely to have a large number of potential matches, and our candidate generation step is likely to have surfaced many of these potentially relevant snippets.

\subsection{LLM-Human Agreement}
\label{sec:agreement}

After automated labelling, we had three individual annotators review model annotations, one of them being a music professional and leading the reviewing efforts. For this step, we target a confidence level of 95\% (i.e., $Z = 1.96$) and a margin of error of $e = 0.05$. Following the standard sample-size estimate for a proportion,
\begin{equation}
n = \frac{Z^2\,p(1-p)}{e^2},
\end{equation}
we use the conservative choice $p = 0.5$, maximizing variance, yielding
\begin{equation}
n = \frac{(1.96)^2\cdot 0.5\cdot (1-0.5)}{(0.05)^2} \approx 384.16 \approx 385.
\end{equation}
This results in us sampling 385 annotated query<->song pairs for human review.

For human verification, we sampled 385 query–song pairs using a stratified scheme over relevance labels (0–3) and query styles to avoid class imbalance. Three annotators (including one professional musician) independently assessed each pair by listening to the audio and reading the query, blind to the model-assigned label and without access to candidate lists or song cards. Disagreements were resolved through adjudication discussions to reach a final consensus label. Agreement after adjudication was unanimous with no reviewer reporting concerns. 

In Table~\ref{tab:agreement}, we present per-label and macro-averaged precision metrics to showcase general alignment, as well as an overview of the re-labelling suggested by annotators as well as the quadratic weighted Cohen's $\kappa$ between human consensus and LLM annotations. Per-label precision is computed as the proportion of model-assigned labels matching the final human consensus.

These results appear to show that the annotations generated by Gemini 3 Pro are highly aligned with those of expert human annotators, with the most apparent issue being a tendency towards leniency at the lower end of the scale, the most common mistake being irrelevant tracks, which should have been labeled ``0'' being labeled as tangentially relevant (``1'') instead.

We also observed that errors appeared to be uniformly distributed, with no statistically significant degradation tied to any of the attributes we control for, such as query length, negation, or number or types of targeted attributes.

\subsection{IncompeBench-Strict and IncompeBench-Lenient}
\label{sec:strict}

Our analysis of the assigned labels reveals that the vast majority of queries have multiple strong matches in the data, but also that the majority of model mistakes are related to over-leniency on the "tangentially relevant" (``1'') label. Building on this insight, we choose to introduce two evaluation sets: \textbf{IncompeBench-lenient}, where qrels are provided with 3 levels of relevance, matching our rating system's original labels, and \textbf{IncompeBench-strict}, where ``1'' annotations are fully discarded and only snippets rated ``2'' or ``3'' are kept as annotations, with ``3'' indicating stronger matches.

\section{Baseline Evaluations}

\begin{table*}[!htbp]
\centering
\small
\setlength{\tabcolsep}{4.5pt}
\begin{tabular}{lcccccccccccc}
  \toprule
  & \multicolumn{3}{c}{nDCG} & \multicolumn{3}{c}{MAP} & \multicolumn{3}{c}{Recall} & \multicolumn{3}{c}{Precision} \\
  \cmidrule(lr){2-4}\cmidrule(lr){5-7}\cmidrule(lr){8-10}\cmidrule(lr){11-13}
  Model & @10 & @50 & @100 & @10 & @50 & @100 & @10 & @50 & @100 & @10 & @50 & @100 \\
  \midrule
  \multicolumn{13}{c}{\textbf{IncompeBench-Strict}}\\
  \midrule
  LAION-CLAP~\cite{clap}      & 0.4596 & 0.4154 & 0.3960 & 0.0397 & 0.1341 & 0.1922 & 0.0501 & 0.1938 & 0.3085 & 0.5990 & 0.4810 & 0.3843 \\
  TTMR++~\cite{ttmr}          & 0.5005 & 0.4603 & 0.4434 & 0.0449 & 0.1518 & 0.2200 & 0.0557 & 0.2175 & 0.3453 & 0.6552 & 0.5335 & 0.4302 \\
  ColQwen-Omni~\cite{colpali} & 0.5129 & 0.5020 & 0.4841 & 0.0464 & 0.1773 & 0.2590 & 0.0576 & 0.2472 & 0.3869 & 0.6714 & 0.5996 & 0.4791 \\
  CLAMP3~\cite{clamp3}        & \textbf{0.5699} & \textbf{0.5368} & \textbf{0.5606} & \textbf{0.0492} & \textbf{0.1838} & \textbf{0.3134} & \textbf{0.0591} & \textbf{0.2494} & \textbf{0.4527} & \textbf{0.7151} & \textbf{0.6370} & \textbf{0.5906} \\
  \midrule\midrule
  \multicolumn{13}{c}{\textbf{IncompeBench-Lenient}}\\
  \midrule
  LAION-CLAP~\cite{clap}      & 0.5942 & 0.5351 & 0.4829 & 0.0396 & 0.1542 & 0.2233 & 0.0423 & 0.1733 & 0.2742 & 0.8570 & 0.6984 & 0.5492 \\
  TTMR++~\cite{ttmr}          & 0.6427 & 0.5847 & 0.5338 & 0.0437 & 0.1694 & 0.2508 & 0.0455 & 0.1883 & 0.3013 & 0.9148 & 0.7568 & 0.6044 \\
  ColQwen-Omni~\cite{colpali} & 0.6534 & 0.6427 & 0.5877 & 0.0443 & 0.2023 & 0.3033 & 0.0462 & 0.2162 & 0.3417 & 0.9280 & 0.8697 & 0.6866 \\
  CLAMP3~\cite{clamp3}        & \textbf{0.6981} & \textbf{0.6804} & \textbf{0.7027} & \textbf{0.0451} & \textbf{0.2129} & \textbf{0.4105} & \textbf{0.0466} & \textbf{0.2253} & \textbf{0.4404} & \textbf{0.9461} & \textbf{0.9204} & \textbf{0.9027} \\
  \bottomrule
\end{tabular}
\caption{Baseline performance on IncompeBench under \textbf{strict} (scores 2--3 positive) and \textbf{lenient} (scores 1--3 positive) relevance settings.}
\label{tab:results}
\end{table*}

We report baseline evaluations for common music retrieval models with publicly available weights: CLAP~\cite{clap}, TTMR++~\cite{ttmr}, CLAMP3~\cite{clamp3}, the current state-of-the-art model across existing tasks, and ColQwen-Omni~\cite{colqwenomni}, a novel attempt to port the ColPali~\cite{colpali} image retrieval recipe to audio through transfer learning on an omnimodal base model. 

We report three metrics: nDCG, the core metric, as well as MAP, Recall and Precision for thoroughness. We compute nDCG using gains equal to the graded relevance (0–3, with 1 dropped for the \emph{Strict} setting), and binarize relevance for Recall/MAP under the strict (2–3) and lenient (1–3) settings.

Table~\ref{tab:results} presents the results for both IncompeBench-Strict and IncompeBench-Lenient. Overall, we observe that there are significant variations between the best and worst performing models.

An interesting phenomenon, common to all models evaluated, is significantly stronger performance on IncompeBench-Lenient than on IncompeBench-Strict. While somewhat expected, due to the large number of ``tangentially relevant'' annotations in the data, these results appear to suggest that rather than surfacing completely unrelated songs, most current models have a strong capability to reliably retrieve at least tangentially relevant results. This is further confirmed by the Precision indicators: in the lenient settings, all three baselines score above 0.9. However, all models appear to struggle noticeably more to capture all the nuances of a given query, as indicated by the both the nDCG measurements and the noticeable drop in precision between the Lenient and Strict settings, showing that tangentially relevant results frequently get ranked higher than songs that better encompass these nuances

We believe that these results both support our theory that the benchmarks provide strong signal towards improving the evaluation of music retrieval models and suggest that there is still significant room for improvement in fine-grained music ranking, which is an increasingly substantial component of many real-world multimodal systems.

\section{Conclusion}

In this work, we introduce IncompeBench, a fine-grained music retrieval benchmark comprising over 125,000 auto-generated annotations strongly aligned with expert human judgements. This large number of multi-level relevance annotations allows IncompeBench to capture nuances that are inherent to music retrieval, where dozens or hundreds of individual songs may be relevant to a given query. To the best of our knowledge, IncompeBench is the first publicly available music retrieval benchmark of its kind encompassing this level of information, leveraging high-quality permissively licensed music. By providing two evaluation variants, IncompeBench-Strict and IncompeBench-Lenient, we enable researchers to assess both coarse and fine-grained ranking quality, with our baseline results demonstrating that current models struggle in particular with the latter, reliably surfacing tangentially relevant results but failing to capture the full nuance of complex queries. The permissive licensing of both the underlying audio corpus and all benchmark artifacts ensures that IncompeBench can be freely redistributed and extended, removing a key barrier that has historically limited reproducibility in music retrieval research. We hope that IncompeBench will serve as a foundation for driving progress in text-to-music retrieval, and we publicly release all data, queries, annotations, and generation code to support this goal.

\section*{Acknowledgements}

We extend our thanks to Kevin MacLeod, the composer of all IncompeTech music, which was used in this benchmark, both for his work in creating the most widely used collection of permissively licensed music and for his supportive words when learning of our efforts towards improved music information retrieval.

\section*{Limitations}

We identify two main limitations to our work. The first is that the songs, while high quality and diverse in genre and instruments, are largely instrumentals from a single prolific composer. Future work should seek to extend this benchmark to cover music containing vocals and a larger volume of sources. As it stands, IncompeTech's wide dissemination and varied catalogue makes it one of the best sources of permissively licensed diverse music tracks without requiring significant scraping and quality filtering efforts. As such, for the purpose of this benchmark and with limited existing, high-quality, readily available resources, we believe that emphasizing musical diversity over authorship diversity is beneficial to the benchmark's quality.

The second limitation is that, although our annotation efforts were substantial, not every pair of the dataset has been annotated and many queries are purposefully broad, therefore the risk of false negative therefore remains. However, this factor is inherent to many, if not all, information retrieval benchmarks~\cite{evals}, thus further reinforcing the need for diverse benchmarking practices.

\bibliographystyle{ACM-Reference-Format}
\bibliography{sample-base}

\appendix

\end{document}